\documentclass[prl,aps,twocolumn,groupedaddress,floats,showpacs,final,superscriptaddress]{revtex4-1}
\usepackage[latin3]{inputenc}
\usepackage[makeroom]{cancel}
\usepackage{graphicx}
\usepackage{amsmath}
\usepackage{amsfonts}
\usepackage{amssymb}
\usepackage{chngcntr}
\usepackage{color}
\usepackage[left=2cm,right=2cm,top=2cm,bottom=2cm]{geometry}

\usepackage{graphicx}
\usepackage{dcolumn}
\usepackage{bm}
\usepackage{simplewick}
\usepackage{array}
\usepackage{appendix}

\RequirePackage[
   hyperindex,colorlinks,bookmarksnumbered,
   plainpages=true,pdfstartview=FitH]{hyperref}
\hypersetup{linkcolor=blue,urlcolor=blue,citecolor=blue}
\usepackage{hyperref}

\definecolor{purple}{rgb}{0.5,0,0.6}

\begin{document}

\title{Ratchet Hall effect in fluctuating superconductors}

\author{A.~V.~Parafilo}
\email[Corresponding author: ]{aparafil@ibs.re.kr}
\affiliation{Center for Theoretical Physics of Complex Systems, Institute for Basic Science (IBS), Daejeon 34126, Korea}

\author{V.~M.~Kovalev}
\affiliation{Novosibirsk State Technical University, Novosibirsk 630073, Russia}

\author{I.~G.~Savenko}
\email[Corresponding author: ]{ivan.g.savenko@gmail.com}
\affiliation{Department of Physics, Guangdong Technion--Israel Institute of Technology, 241 Daxue Road, Shantou, Guangdong, China, 515063}
\affiliation{Technion -- Israel Institute of Technology, 32000 Haifa, Israel}

\date{\today}

\begin{abstract}
We propose a superconducting ratchet-induced Hall effect (RHE), characterized by the emergence of a unidirectional, rectified flux of fluctuating Cooper pairs in a two-dimensional thin film exposed to an external electromagnetic field. 
The RHE is a second-order response with respect to the electromagnetic field amplitude. 
It consists of a nonzero photocurrent due to the breaking of the system's inversion symmetry driven by the combined action of the in-plane time-dependent electric field and a spatial modulation of the critical temperature. 
We explore a means to control the electric current by the polarization of the external field accompanied by a non-linear Hall effect of fluctuating Cooper pairs caused by circularly polarized irradiation. 
Moreover, the nonlinear conductivity tensor exhibits a higher-power dependence on the reduced temperature compared to that of the conventional Aslamazov-Larkin correction (or other fluctuating second-order nonlinear responses). 
It results in a dramatic enhancement of the non-linear Hall response of fluctuating Cooper pairs in the vicinity of the superconducting criticality.
\end{abstract}

\maketitle

The Hall effect consists in the emergence of a transverse voltage or electric current in the sample in response to a longitudinal field $j_\alpha=\sigma_{\alpha\beta}E_\beta$ (where $\sigma_{\alpha\beta}$ is the conductivity and $\mathbf{E}$ is the electric field) under certain constraints.
One of them is usually a broken time-reversal symmetry, which is conventionally provided by an external magnetic field normal to the plane of the sample. 
However, the effect may also take place without the magnetic field (if some magnetic order is present)~\cite{RevModPhys.82.1539, PhysRev.37.405}. 
The Hall effects without the presence of an external magnetic field are usually called anomalous Hall effects.
Generally, broken symmetries in the crystal
lattice or in the structure itself lead to specific features
of electron-photon~\cite{RefBelinisher, PismaZhETF.27.640, Ganichev_2003, GLAZOV2014101}, electron-phonon~\cite{Kibis_1999, Kibis_2001}, and electron-electron
interactions~\cite{PhysRevB.106.235305, RefMineev, PhysRevB.110.L041301, PhysRevB.109.245414, Kibis_1992, Kibis_2002} in a wide range of low-dimensional
systems.
One of the intrinsic mechanisms providing the Hall current is the Berry phase~\cite{RevModPhys.82.1959}.
Furthermore, without the breaking time-reversal symmetry, the Hall effect can emerge utilizing the spin~\cite{RevModPhys.87.1213} or valley~\cite{PhysRevLett.99.236809, PhysRevB.77.235406, PhysRevLett.108.196802, doi:10.1126/science.1250140} degrees of freedom or outside of the linear-response regime. 

The nonlinear Hall effect (NHE)~\cite{Zhang_2018, PhysRevLett.121.266601, RefMa, RefKang} consists in a transverse current in response to two longitudinal driving fields (e.g., the second power of the electric field or even a mixture of electric and magnetic fields): $j_\alpha=\chi_{\alpha\beta\gamma}E_\beta E_\gamma$~\cite{Du2021PerspectiveNH}. 
It does not require the breaking of time-reversal symmetry, instead, the inversion symmetry is to be violated. 
Consequently, NHE allows for creating powerful tools to explore the symmetry and topology of novel materials and quantum matter in general.
%
It represents a nonlinear nonreciprocal quantum transport phenomenon, which by its essence requires lowered system symmetry~\cite{RefDu2021}. 
Thus, it is expected to take place in three-dimensional Weyl semimetals and two-dimensional (2D) Dirac materials such as transition metal dichalcogenides~\cite{RefMa, RefKang}, and in topological insulators~\cite{PhysRevLett.115.216806}.

The study of the non-linear response, including the NHE, in superconducting (SC) materials has recently gained considerable attention, with research focus both below and above the critical temperature $T_c$. 
Such a growing interest underscores the importance of our understanding of the nonlinear phenomena for the purpose of basic science and for advancing SC technologies. In particular, recent studies demonstrate the significance of the inversion and time-reversal symmetries~\cite{PhysRevB.100.220501,PhysRevB.108.224516, PhysRevB.105.024308}, gauge invariance~\cite{PhysRevB.106.094505,watanabe}, and the Berry curvature factor in SC systems~\cite{PhysRevB.107.024513} in non-linear second-order response, and the possibility to design photoinduced supercurrent-based devices~\cite{PhysRevB.108.L180509,Parafilo_2025}.  
Such phenomena serve as valuable tools for exploring the symmetry of SC pairing, paving the way to manipulate the flow of the SC condensate using optical-only methods.

As one of the directions of research, one should not overlook the importance of fluctuating superconductors, when the system is in the normal state but near the superconducting criticality, leading to the formation of Cooper pairs with a finite lifetime that remain resilient against disorder. 
The sensitivity of fluctuating Cooper pairs to temperature can significantly enhance the nonlinear response as compared with the contribution of single electrons~\cite{PhysRevB.103.024513,PhysRevB.106.144502,boev_kovalev,PhysRevB.111.155120,PhysRevResearch.6.L022009}. 


In this Letter, we explore the Ratchet Hall effect (RHE) as an alternative to the nonlinear photoresponse in fluctuating superconductors. 
The term `ratchet' in condensed matter  refers to the transport phenomena occurring in systems with a potential and a broken spatial symmetry
far from thermal equilibrium (even in the absence of an average macroscopic force)~\cite{REIMANN200257, RevModPhys.81.387}. 
There exist many models that describe the ratchet effect. 
We will follow the \textit{Seebeck ratchet} mechanism: a model in which the non-equilibrium regime is achieved through a spatially dependent periodic temperature profile $T(x)$, which together with the spatial potential $V(x)$ provides a directional transport due to a non-zero spatial average $\overline{T(x)dV(x)/dx}$.
A particular realization of a temperature Seebeck ratchet proposed in Ref.~\cite{PhysRevLett.81.4040} has been recently realized experimentally in semiconductor quantum well-based structures~\cite{PhysRevLett.103.090603,ratchet_JETFLetter,PhysRevB.83.165320}.
We will show that the Seebeck ratchet can be implemented in fluctuating superconductors, resulting in a nontrivial Hall transport of fluctuating Cooper pairs.




%
%
%
\begin{figure}[t!]
\includegraphics[width=1\columnwidth]{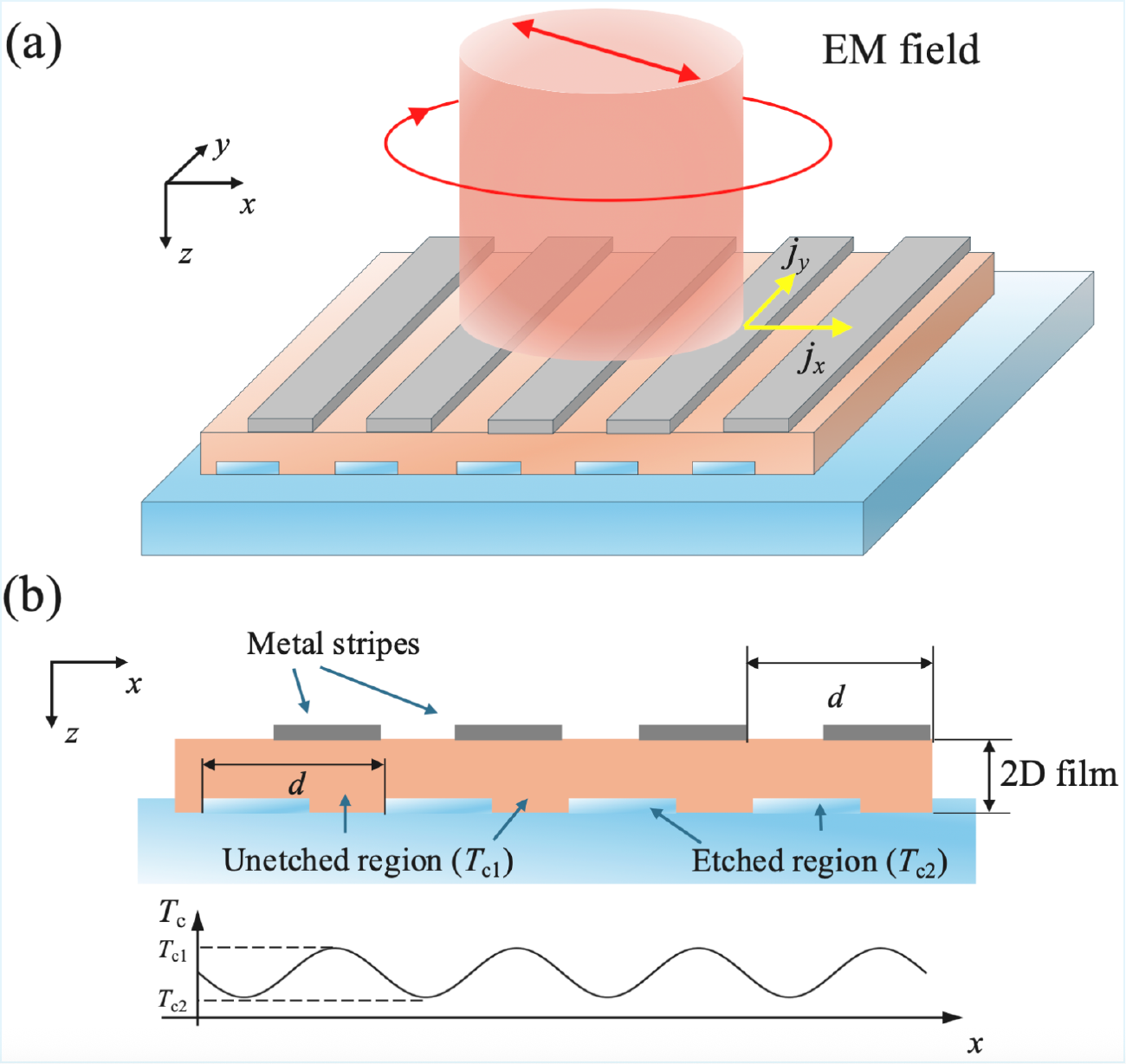}
\caption{System schematic: (a) A superconducting thin film with modulated critical temperature in a normal state ($T\gtrsim T_c(x)$) exposed to an external electromagnetic field with circular (or linear) polarization, whose in-plane projection of the electric field is modulated along $x$ direction. 
The modulation of the time-dependent electric field is induced by a special mask, which consists in metallic strips with the period $d$. 
(b) Profile of the system together with the spatial dependence of the critical temperature. 
The critical temperature of the etched regions $T_{c2}$ is lower than the critical temperature of the unetched regions $T_{c1}$.}
\label{Fig1}
\end{figure}
%
%
%



\textit{Theoretical framework.---}Consider a 2D SC film at the temperature $T\gtrsim T_c$ exposed to a circularly or linearly polarized electromagnetic (EM) field with a frequency $\omega$. 
The SC critical temperature is spatially modulated with the period $d$ along the $\textbf{e}_x$ axis (Fig.~\ref{Fig1}). 
Thus, the system represents a thin-film analog of a $T_c$-modulated superlattice. 
In an experiment, such a $T_c$ modulation can be achieved by a reactive-ion etching~\cite{PhysRevLett.65.2905,PhysRevB.45.9850} (see also our comment~\cite{[{Alternatively, a similar spatial modulation of $T_{c}$ can be realized in a 2D film attached to a massive superconductor. 
As is known, induced by the proximity effect superconductivity $\Delta_{in}$ is determined by the properties of a massive superconductor and an insulating layer, $\Delta_{in}\sim |t|^2\Delta$ with $\Delta$ the order parameter of the massive superconductor and $t$ the tunnel hopping of the insulating layer. Assuming spatial modulation of the insulating layer' thickness, one can define $t\rightarrow t(x)$ to account for this variation. As a consequence, $\Delta_{in}=\Delta_{in}(x)$ and $T_c=T_c(x)$. Yet another possible mechanics is based on the periodic modulation of magnetic-impurity concentration in the 2D SC film studied in Ref.~\cite{kulik}}]Comment}). 
The critical temperature of the etched regions $T_{c2}$ is lower than the one of the unetched regions $T_{c1}$. 
Furthermore, the time-dependent EM field amplitude is also weakly spatially modulated along the $x$ direction (with the same period as the critical temperature modulation), which is achieved by the deposition of metallic stripes along the SC film:  
%
\begin{eqnarray}\label{temp_modul}
T_c(\textbf{r})= T_{c0} + \delta T_c \cos(\textbf{q}\cdot\textbf{r}),
\end{eqnarray}
\begin{eqnarray}
\textbf{E}(\textbf{r},t)=(\textbf{E}e^{-i\omega t}+\textbf{E}^{\ast}e^{i \omega t})[1+h\cos(\textbf{q}\cdot\textbf{r}+\phi)],
\end{eqnarray}
where $T_{c0}=(T_{c1}+T_{c2})/2$, $\delta T_{c}=(T_{c1}-T_{c2})/2$; $\textbf{q}=(q,0)$ with $q=2\pi/d$, and $h\ll1$ is a parameter characterizing the weakness of the EM field modulation induced by the metallic stripes, and $\phi$ describes the phase shift between the temperature and EM field modulations. 
The amplitudes of the linearly and circularly polarized EM wave read $\textbf{E}=(E_x,E_y)$ and $\textbf{E}=(1,i\sigma)E_0$, respectively, with $\sigma=\pm 1$ indicating the polarization.
By the normal phase we mean $T > T_{c0}+\delta T_c$ in what follows to avoid the emergence of any SC islands. 
Also note that we only examine the case of a normally illuminated EM wave that is propagating perpendicular to the plane of the SC thin film.

The stationary second-order photo response of fluctuating Cooper pairs consists in the emergence of the current density, which is composed of the paramagnetic and diamagnetic parts, $\textbf{j}=\textbf{j}_P+\textbf{j}_D$:
\begin{subequations}
\begin{align}
&\textbf{j}_P=\frac{e}{2m}\left(\langle\overline{\Psi^{\ast}(\textbf{r}',t')\hat{\textbf{p}}\Psi(\textbf{r},t)}\rangle-\langle\overline{\Psi(\textbf{r}',t')\hat{\textbf{p}}\Psi^{\ast}(\textbf{r},t)}\rangle\right),\\
&\textbf{j}_D=-\frac{2e^2}{m}\overline{\textbf{A}(\textbf{r},t)\langle\Psi^{\ast}(\textbf{r}',t')\Psi(\textbf{r},t)\rangle},
\end{align}
\end{subequations}
where $m$ is the particles' effective mass, $\hat{\textbf{p}}=-i\nabla$ is a momentum operator, $\mathbf{A}$ is a vector potential of the EM field, $\Psi(\textbf{r},t)$ is the order parameter, and the bar indicates averaging along the $x$ axis.

We describe the system using the time-dependent Ginzburg-Landau (TDGL) equation for the order parameter 
(now using $\hbar=k_B=c=1$ units):
\begin{eqnarray}
\nonumber
&&\left\{\gamma \left(\frac{\partial}{\partial t}+2ie\varphi(\textbf{r},t)\right) 
+\alpha T_{c0} \left(\epsilon_0+\xi^2[\hat{\textbf{p}}-2e\textbf{A}(\textbf{r},t)]^2\right)
\right.\\
\label{TDGL}
&&~~~~~~~~~~~~~~\left.
-\alpha \delta T_c \cos(\textbf{q}\cdot\textbf{r})\right\}\Psi(\textbf{r},t)
=f(\textbf{r},t).
\end{eqnarray}
Here, the dimensionless coefficient $\gamma$ consists of the real and imaginary parts: $\gamma=\gamma'+i\gamma''$, where  
\begin{eqnarray}\label{gammaparameter}
\gamma'=\frac{\pi \alpha}{8}=\alpha T_{c0}\epsilon_0\tau_{GL}\quad, \quad\gamma''=-\frac{\alpha T_{c0}}{2}\frac{\partial \log(T_{c0})}{\partial E_F}.
\end{eqnarray}
$\alpha =1/4mT_{c0}\xi^2$ is the parameter of the Ginzburg-Landau theory, $\xi$ the coherence length, and $\epsilon_0=\log(T/T_{c0})\approx (T-T_{c0})/T_{c0}$ is a reduced temperature.

The real part of $\gamma$ determines the lifetime of fluctuating Cooper pairs, $\textrm{Re}[\gamma]\propto \tau_{GL}\equiv (\pi/8)(T-T_{c0})^{-1}$, while the imaginary part appears as a consequence of electron-hole asymmetry in the band structure or topological features of the Fermi surface~\cite{Varlamov}. 
It should be noted, that $\gamma''$ may play significant role in a number of effects, resulting, e.g., in a nonzero fluctuating Hall effect~\cite{PhysRevB.51.3880} or nonzero photon drag of fluctuating Cooper pairs~\cite{PhysRevB.101.104512}. 
 
The coherence length in a 2D superconductor reads~\cite{Varlamov}
\begin{eqnarray}\label{coherence_length}
\xi^2=\frac{v_F^2\tau^2}{2}\Bigl[
\psi\left(\frac{1}{2}\right)
-
\psi\left(\frac{1}{2}+\frac{1}{4\pi T\tau}\right)
+\frac{\psi'\left(\frac{1}{2}\right)}{4\pi T\tau}\Bigr],~~~
\end{eqnarray}
where $\psi(x)$ is the digamma function, $v_F$ is the Fermi velocity, and $\tau$ is a relaxation time. 
In the limiting cases, Eq.~\eqref{coherence_length} can be simplified: $\xi^2_d\approx(\pi v_F^2/8T)(\tau /2)$ in dirty samples ($\tau T\ll1$), and $\xi^2_c\approx\pi v_F^2/16T^2$ in  clean films ($\tau T\gg1$). 
In our case of a spatially modulated sample, 
using Eq.~(\ref{TDGL}) is justified if the period of spatial modulations is larger than the coherence length: $\xi\ll d$.

The electrostatic and vector potentials read
\begin{eqnarray}
\varphi(\textbf{r},t)=(\varphi e^{i\textbf{k}\cdot\textbf{r}-i\omega t}+ \varphi^{\ast} e^{-i\textbf{k}\cdot\textbf{r}+i\omega t})\\
\nonumber\times[1+h\cos(\textbf{q}\cdot\textbf{r}+\phi)],
\end{eqnarray}
\begin{subequations}
    \begin{align}
        &\textbf{A}(\textbf{r},t)=\textbf{A}(t)[1+h\cos(\textbf{q}\cdot\textbf{r}+\phi)],\\
&\textbf{A}(t)=\textbf{A}e^{-i\omega t}+\textbf{A}^{\ast}e^{i\omega t}.
    \end{align}
\end{subequations}
%
Note, that the length gauge, when the electric field is introduced via the scalar potential, is more convenient in the case of linear polarization. 
In contrast, the use of the vector potential (velocity gauge) is more convenient for the circularly polarized EM waves.  

The right-hand side of Eq.~(\ref{TDGL}) is the Langevin force, describing SC fluctuations in the equilibrium. It satisfies the Gaussian
white-noise law:
\begin{eqnarray}\label{white_noise}
\langle f(\textbf{r},t) f^{\ast}(\textbf{r}',t')\rangle =2 \gamma' T \delta(\textbf{r}-\textbf{r}')\delta(t-t'),
\end{eqnarray}
where $\langle ... \rangle$ stands for the averaging over fluctuations.
%
Considering the scalar and vector potentials and the critical temperature modulation as  perturbations, we find the solution of Eq.~(\ref{TDGL}) by expanding the order parameter up to the third order: 
\begin{eqnarray}\label{OP}
\Psi=\Psi_0+\Psi_1+\Psi_2+\Psi_3+O(\Psi_4),
\end{eqnarray}
where $\Psi_0$ is a fluctuating order parameter in the equilibrium, $\Psi_1\propto \{\textbf{E},\delta T_c\}$, $\Psi_2\propto \{\textbf{E}^2,\textbf{E}\,\delta T_c\}$, and $\Psi_3 \propto \textbf{E}^2\delta T_c$. 
The third-order correction gives the highest non-vanishing contribution because $\textbf{j}=0$ in the absence of spatial modulation ($h=0$ or $\delta T_c=0$). 
Indeed, accounting for the spatial modulation of the EM field ($h\neq 0$) only is not sufficient since $\textbf{j}\propto h \textbf{E}^2$ vanishes after averaging over $x$ direction, 
and to achieve a nontrivial photoresponse, one must break the spatial symmetry. 
The only way to accomplish this in current setup is to introduce an additional spatial modulation associated with the critical temperature $\delta T_c\neq0$. 
Thus, we envisage the effect $|\textbf{j}|\sim h q|\textbf{E}|^2 \delta T_c$, where the parameter $q$ should emerge after the averaging along $x$ direction. 

The order parameter $\Psi(\textbf{r},t)$ can be found by iterations after substituting the expansion~(\ref{OP}) in Eq.~(\ref{TDGL}), yielding
\begin{subequations}\label{OP2}
    \begin{align}
        &\Psi_0(\textbf{r},t)= \int d\textbf{r}_1dt_1 \hat{G}(\textbf{rr}_1;tt_1) f(\textbf{r}_1,t_1),\\
&\Psi_n(\textbf{r},t)=\int d\textbf{r}_1dt_1 \hat{G}(\textbf{rr}_1;tt_1)\hat W_{\nu}(\textbf{r}_1,t_1)\Psi_{n-1}(\textbf{r}_1,t_1),\label{correction}
    \end{align}
\end{subequations}
where $\hat{G}(\textbf{rr}';tt')=\sum_{\varepsilon\textbf{p}} \hat{G}_{\textbf{p}}(\varepsilon) e^{i\textbf{p}\cdot(\textbf{r}-\textbf{r}')-i\varepsilon (t-t')}$ is a standard propagator of fluctuating Cooper pairs~\cite{Varlamov}, whose Fourier transform reads as
\begin{eqnarray}
\hat{G}_{\textbf{p}}(\varepsilon)=\frac{1}{-i\gamma \varepsilon +\alpha T_{c0}(\epsilon_0+\xi^2\textbf{p}^2)}.
\end{eqnarray}
In Eq.~(\ref{correction}), $\hat{W}_{\nu}$ is a perturbation for the case of linear ($\nu="\textrm{lp}"$) and circular ($\nu="{\rm cp}"$) polarization of the EM field:
\begin{subequations}\label{perturbation}
    \begin{align}\label{lin}
        &\hat{W}_{\rm lp}(\textbf{r},t)= -2ie\gamma \varphi(\textbf{r},t)+\alpha \delta T_c \cos(\textbf{q}\cdot\mathbf{r}),\\
&\hat W_{\rm cp}(\textbf{r},t)=\frac{e}{m}\textbf{A}(\textbf{r},t)\cdot\hat{\textbf{p}}+\frac{ieh}{2m}\textbf{q}\cdot\textbf{A}(t)\sin(\textbf{q}\cdot\mathbf{r}+\phi)\nonumber\\
&~~~~~~~~~~~~~~~~+\alpha\delta T_c\cos(\textbf{q}\cdot\mathbf{r}).\label{circ}
    \end{align}
\end{subequations}
%

Furthermore, we can distinguish the following essential contributions to the paramagnetic,
\begin{subequations}\label{paramag}
\begin{align}
&\textbf{j}_P^{I}=\frac{e}{2m}\left(
\overline{\langle\Psi_0^{\ast}\hat{\textbf{p}}\Psi_3\rangle}+\overline{\langle\Psi_3^{\ast}\hat{\textbf{p}}\Psi_0\rangle}+\textrm{c.c.}\right),\label{paramagnetic1}\\
&\textbf{j}_P^{II}=\frac{e}{2m}\left(
\overline{\langle\Psi_1^{\ast}\hat{\textbf{p}}\Psi_2\rangle}+\overline{\langle\Psi_2^{\ast}\hat{\textbf{p}}\Psi_1\rangle}+\textrm{c.c.}\right),\label{paramagnetic2}
\end{align}
\end{subequations}
and diamagnetic, 
\begin{subequations}\label{diamag}
\begin{align}
&\textbf{j}_D^{I}=-\frac{2e^2}{m}\overline{\textbf{A}(\textbf{r},t)\left(\langle\Psi_0^{\ast}\Psi_2\rangle+\langle\Psi_2^{\ast}\Psi_0\rangle\right)},\\
&\textbf{j}_D^{II}=-\frac{2e^2}{m}\overline{\textbf{A}(\textbf{r},t)\langle\Psi_1^{\ast}\Psi_1\rangle},
\end{align}
\end{subequations}
current densities. 
Let us mention that in this manuscript, we are only interested in the static nonlinear photo-response, thus disregarding the effects of second-harmonic generation.



\textit{Results and discussion.---}First, we consider the case of circularly polarized EM field: $\textbf{A}=(1,i\sigma)A_0$, $\textbf{A}=\textbf{E}/i\omega$. 
The calculations 
show that, in this case, the main contribution is the diamagnetic current~(\ref{diamag}), which provides the following expression for the ratchet Hall current density:
\begin{eqnarray}
\label{current_circ}
&&j_y=-\sigma\frac{4e^3}{\pi}\frac{\gamma''}{\gamma'}hq\xi^2\sin(\phi) \frac{\tau_{GL}^2}{\tilde\omega^2} \frac{T\delta T_c}{T_{c0}\epsilon_0^2} \mathcal{F}_{\rm cp}(\tilde\omega)E_0^2,\\
\label{function_circ}
&&\mathcal{F}_{\rm cp}(\tilde\omega)=\frac{1}{\tilde\omega^2}\left[2\pi -4\arctan\left(\frac{2}{\tilde\omega}\right)-\tilde\omega-\frac{2\tilde\omega}{4+\tilde\omega^2}\right],~
\end{eqnarray}
where $\tilde\omega\equiv \omega \tau_{GL}=\pi \omega/8(T-T_{c0})$, and $\mathcal{F}_{\rm cp}$ is a nonmonotonous function of $\tilde\omega$ which changes its sign and has the following asymptotics at small and large frequencies: $\mathcal{F}_{\rm cp}(\tilde\omega\ll1)\approx (\tilde\omega/12)(1-9\tilde\omega^2/20)$ and $\mathcal{F}_{\rm cp}(\tilde\omega\gg1)\approx -1/\tilde\omega$, respectively.

\begin{figure}[t!]
\includegraphics[width=1\columnwidth]{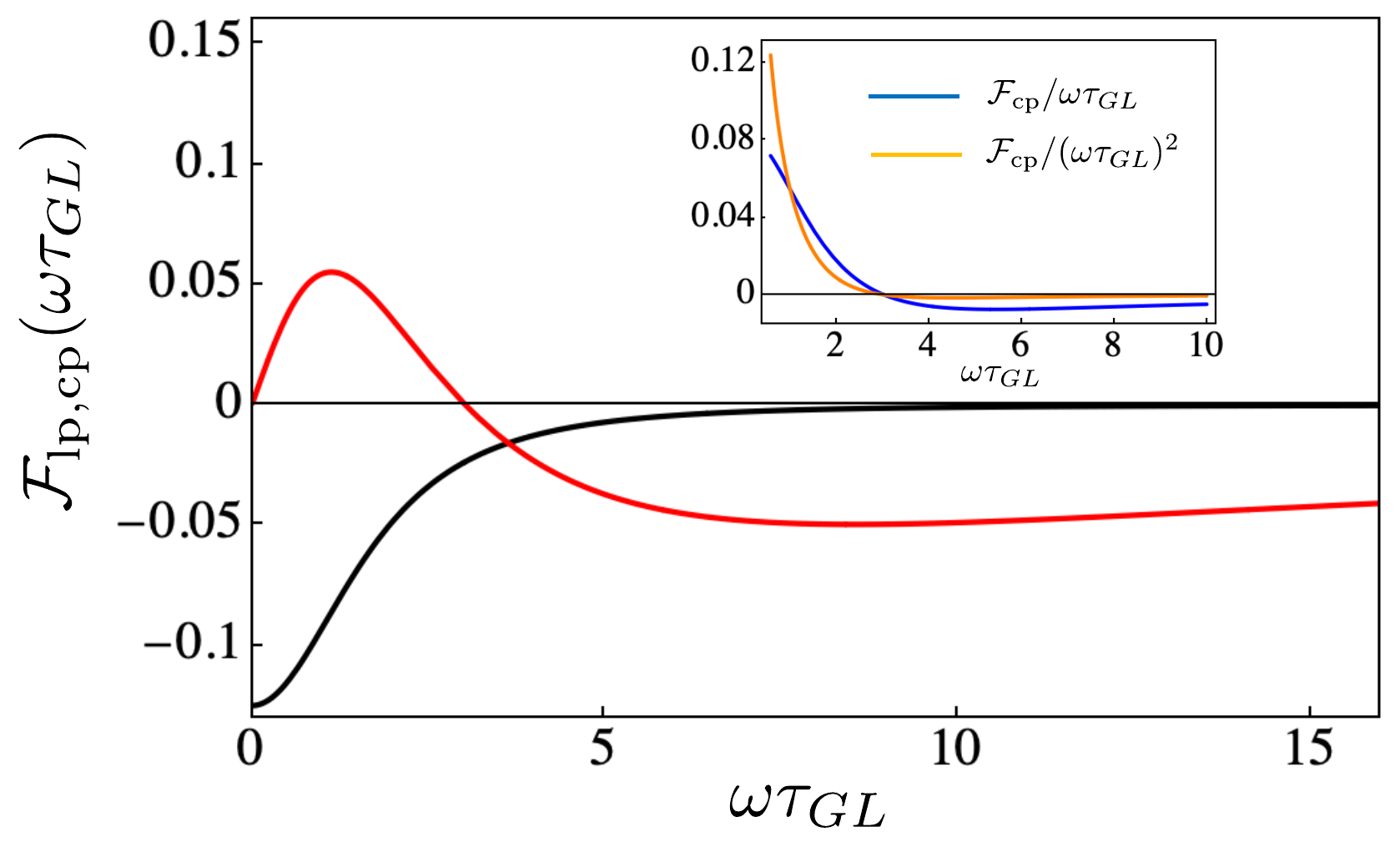}
\caption{Normalized spectra of the Hall electric current density in the case of circular (red) and linear (black) polarization of the EM field: The dependencies of the dimensionless functions $\mathcal{F}_{\rm cp}$ from Eq.~(\ref{current_circ}) (red) and $\mathcal{F}_{\rm ln}$ from Eq.~(\ref{function_circ}) (black) on the dimensioless frequency $\tilde\omega\equiv \omega\tau_{GL}=\pi\omega/8(T-T_{c0})$. Inset: Functions $\mathcal{F}_{\rm cp}(\tilde\omega)/\tilde\omega$ and $\mathcal{F}_{\rm cp}(\tilde\omega)/\tilde\omega^2$. 
}
\label{Fig2}
\end{figure}

The ratchet current for the linearly polarized EM field is based on Eq.~(\ref{lin})~\cite{[{The preference for the length gauge while calculating the ratchet effect induced by the linearly-polarized EM wave is because calculation using the vector-potential can lead to an unphysical result such as quadratic divergency of the current at low frequency, $j_x\propto |\textbf{E}|^2/\omega^2$}]Gauge}. 
The only surviving contribution to the current density stems from the paramagnetic current~(\ref{paramag}).
It should be noted, that to obtain a finite result, one should expand $\overline{\langle \Psi_0^{\ast}\hat{\textbf{p}}\Psi_3\rangle}$, $\overline{\langle \Psi_3^{\ast}\hat{\textbf{p}}\Psi_0\rangle}$, $\overline{\langle \Psi_1^{\ast}\hat{\textbf{p}}\Psi_2\rangle}$ and $\overline{\langle \Psi_2^{\ast}\hat{\textbf{p}}\Psi_1\rangle}$ as well as their complex conjugated counterparts up to the second order in $\textbf{k}$. 
Meanwhile, the related expressions for the zeroth and first-order expansions in $\textbf{k}$ vanish. 
Using the correspondence between the scalar potential and the electric field $-i\textbf{k}\varphi=\textbf{E}$ gives the following expression for the current density:
\begin{eqnarray}
\label{current_lin}
j_x&=&\frac{4e^3}{\pi}\frac{\gamma''}{\gamma'}hq\xi^2\sin(\phi) \tau^2_{GL}\frac{T\delta T_c}{T_{c0}\epsilon_0^2} \mathcal{F}_{\rm lp}(\tilde\omega)\\
\nonumber
&&\times\left(|E_x|^2+|E_y|^2\right),
\end{eqnarray}
and the $y-$projection vanishes.
The analytical form of the function $\mathcal{F}_{\rm lp}(\tilde\omega)$ in~\eqref{current_lin} is too cumbersome to write it here explicitly (see End Matter below). 
We will only mention that $\mathcal{F}_{\rm lp}(\tilde\omega)$ is a monotonous function  with the following asymptotics: $\mathcal{F}_{\rm lp}(\tilde\omega\ll1)\approx (-1/8) (1-\tilde\omega^2/3)$ and $\mathcal{F}_{\rm lp}(\tilde\omega\gg1)\approx 4[1-2\log(\tilde\omega/2)]/\tilde\omega^4$.
Figure~\ref{Fig2} shows the spectra of the longitudinal and transverse current densities.

The general dependence of the ratchet-induced current obeys the following phenomenological expression:
\begin{eqnarray}\label{general}
\textbf{j}=\chi_1 \overline{|\textbf{E}|^2\textbf{F}}+\chi_2 [\overline{\textbf{E}^{\ast}(\textbf{E}\cdot\textbf{F})}+\textrm{c.c}]\\
\nonumber
+i\chi_3 \overline{[\textbf{F}\times [\textbf{E}\times \textbf{E}^{\ast}]]},
\end{eqnarray}
where
$\textbf{F}\propto \nabla T_c(\textbf{r})$, (and the bar, as before, signifies the spatial average). 
The coefficient $\chi_1$ characterizes the EM field-induced correction to isotropic conductivity, while $\chi_2$ and $\chi_3$ describe the anisotropic photoconductivity induced by linearly and circularly polarized EM fields, respectively. 
Finite ratchet current prevails only because the average of two spatial modulations, $\textbf{E}(\textbf{r},t)$ and $\textbf{F}(\textbf{r})$, is nonzero. 
This is a consequence of breaking the spatial invariance (which technically leads to the emergence of $\sin\phi$ factor in Eqs.~(\ref{current_circ}) and (\ref{current_lin})). 
Comparing the terms in Eq.~(\ref{general}) with Eqs.~(\ref{current_circ}) and (\ref{current_lin}), we conclude that in our case $\chi_2=0$, whereas $\chi_1$ and $\chi_3$ are determined by Eqs.~(\ref{current_lin}) and (\ref{current_circ}), respectively. 

Expressions similar to Eq.~(\ref{general}) arise in various problems related to the nonlinear photoinduced phenomena in 2D systems. 
In these problems, $\textbf{F}$ is an in-plane electric field~\cite{PhysRevB.104.085306,boev_kovalev}, an in-plane `photon' momentum~\cite{PhysRevB.101.104512}, or even the velocity of the SC condensate flow~\cite{PhysRevB.108.L180509,Parafilo_2025}.

We conclude that a circularly polarized EM field produces a nonlinear Hall effect of fluctuating Cooper pairs. 
The physical interpretation of this effect is in the inverse Faraday effect: a static magnetization $\textbf{H}_{\rm eff}\propto [\textbf{E}(\omega)\times \textbf{E}^{\ast}(\omega)]$ is established by the circularly polarized light, thus playing a role of an effective magnetic field for the fluctuating Cooper pairs that flow along $\textbf{e}_x \parallel \mathbf{q}$ due to inversion symmetry breaking along that direction. 
Thus, the fluctuating Cooper pairs experience an effective Lorentz force $\mathcal{F}_{LF} \sim 2e[\nabla T_{c}(\textbf{r}) \times \textbf{H}_{\rm eff}]$ in $\textbf{e}_y$ direction. 
This effect is somewhat similar to the account of the Berry curvature and the finite momentum of Cooper pairs, explored in recent studies~\cite{PhysRevB.111.155120, PhysRevResearch.6.L022009}.

Furthermore, both Eqs.~(\ref{current_circ}) and (\ref{current_lin}) demonstrate the properties typical for the `standard' ratchet effect in semiconductor quantum wells with a 1D lateral periodic potential~\cite{PhysRevLett.103.090603, ratchet_JETFLetter, PhysRevB.83.165320}, namely: The ratchet current is $\propto hq$, while its direction is determined by the relative phase between two spatial modulations via $\sin\phi$. 
Furthermore, the coefficient $\chi_1$ remains constant, while $\chi_3(\omega\rightarrow0)\propto \omega^{-1}$, cf. Eq.~(\ref{current_circ}) is formally divergent in the static limit (see the inset in Fig.~\ref{Fig2}). 
The latter statement seems peculiar since one expects $\chi_3(\omega \rightarrow 0) = 0$; 

The issue here is that a circular polarization cannot be adequately described by a static electric field. 
In the conventional ratchet effect in heterostructures~\cite{PhysRevLett.103.090603,ratchet_JETFLetter,PhysRevB.83.165320} and graphene~\cite{PhysRevB.86.115301}, the divergence of the $\chi_3$ coefficient at low frequencies is attributed to the limitations of the perturbation theory. 
In particular, the condition $\tau_{\varepsilon} \ll \tau$, where $\tau_{\varepsilon}$ represents the energy relaxation time and $\tau$ denotes the transport relaxation time, was utilized while deriving the coefficients $\chi_i$ through the Boltzmann kinetic equation. 
The authors in Refs.~\cite{PhysRevLett.103.090603,ratchet_JETFLetter,PhysRevB.83.165320,PhysRevB.86.115301} claim that the limit $\chi_3(\omega\rightarrow0)=0$ is restored at $\omega\sim \tau_{\varepsilon}^{-1}\ll\tau^{-1}$. 
Nevertheless, expression (\ref{current_circ}) appropriately describes the fluctuating ratchet for the terahertz parameter range used in the ratchet experiments~\cite{PhysRevLett.103.090603}. The consideration of the fluctuating ratchet effect in the regime $\omega\ll\tau_{GL}^{-1}$ and a circularly polarized EM field requires a microscopic treatment (and account of the vertex corrections), which we leave beyond the scope of this manuscript.

Furthermore, we want to emphasize that the fluctuating ratchet current is an effect determined by the imaginary part of $\gamma''=\textrm{Im}[\gamma]$ parameter. That puts the ratchet effect on par with such fluctuating effects as the Hall effect, thermoelectric effect, and the photon drag. Although the ratio $\gamma''/\gamma'$ is expectedly small (of the order of $T_{c0}/E_F$), the impact of fluctuations can be significantly enhanced due to the dependence of $\gamma''$ on the topology of the Fermi surface~\cite{Varlamov}. 

Last, but not the least, the coefficients $\chi_1$ and $\chi_3$ from Eqs.~(\ref{current_circ}) and~(\ref{current_lin}) in the low-frequency limit show different power-law dependencies on the Cooper pair lifetime: $\chi_1(\tilde\omega\ll1)\propto \tau_{GL}^2$ and $\chi_3(\tilde\omega\ll1)\propto \tau_{GL}$, respectively. 
Again, it features closely to the normal ratchet effect, where corresponding coefficients have different dependencies on energy and transport relaxation times~\cite{PhysRevLett.103.090603,ratchet_JETFLetter,PhysRevB.83.165320,PhysRevB.86.115301}. 
As the relaxation time in the theory of fluctuating Cooper pairs depends on temperature, it leads to varying reduced temperature dependencies of the ratchet current based on the polarization of the EM field.
Evidently, $\chi_1 \sim \tau^2_{GL}/\epsilon_0^2 \propto (T - T_{c0})^{-4}$ and $\chi_3 \sim \tau_{GL}/\epsilon_0^2 \propto (T - T_{c0})^{-3}$ that grow faster than both the Aslamazov-Larkin $\sigma_{AL}\propto (T-T_{c0})^{-1}$~[\onlinecite{AL}] and the non-linear fluctuating photogalvanic $\sigma_{PVE}\propto (T-T_{c0})^{-2}$~[\onlinecite{PhysRevB.106.144502}] corrections to the Drude conductivity in 2D systems. 

It should be noted, that the applicability of the developed phenomenological theory is primarily determined by the reduced temperature. On one hand, the condition $\epsilon_{0} \approx (T - T_{c0})/T_{c0} \ll 1$ supports the use of the Ginzburg-Landau theory. On the other hand, $\epsilon_{0}$ must also satisfy the Ginzburg-Levanyuk criterion, $\epsilon_{0} \gg \mathrm{Gi}_{(2)} \sim T_{c0}/E_F$, to avoid the temperature range where strong fluctuations occur.


\textit{Conclusions.---}We proposed a theory of a ratchet Hall effect in superconducting films with a spatial modulation of the critical temperature. 
The effect implies the emergence of a unidirectional motion of fluctuating Cooper pairs, which occurs when the film is in a normal state and exposed to an electromagnetic field. 
The ratchet current originates from the breaking of the inversion symmetry in the system, which is influenced by the interaction of two spatial modulations: i) the critical temperature of the film $T_{c}(\textbf{r})\propto \cos(\textbf{qr})$ and ii) the in-plane variation of the electric field component of the EM wave $\textbf{E}(\textbf{r},t)$. 

The direction of the fluctuating ratchet current is sensitive to the polarization of EM field: Linearly polarized light generates the current aligned with the axis that lacks inversion symmetry ($\textbf{j}\parallel\textbf{q}$), while circularly polarized light induces a current in the transverse direction to that axis ($\textbf{j}\perp \textbf{q}$). 
The latter encourages us to assert the existence of the nonlinear Hall effect of fluctuating Cooper pairs, which is induced by a circularly polarized EM field and facilitated by the ratchet mechanism. 
We emphasize that the ratchet-induced nonlinear Hall effect has no single-electron counterpart, while it can be significantly enhanced, $\textbf{j}_{RH}(\tilde\omega\ll1)\sim(T-T_{c0})^{-3}$, near the superconducting criticality.

We believe that the nonlinear ratchet Hall effect proposed in this Letter partially sheds light on the problem of nonlinear photo response in superconductors, thus providing a robust against disorder platform for studying the anomalous Hall effect.


{\it Acknowledgement.} --- We were supported by the Institute for Basic Science in Korea (Project No.~IBS-R024-D1), the National Natural Science Foundation of China, the Natural Science Foundation of Guangdong Province (China), the Ministry of Science and Higher
Education of the Russian Federation (Project FSUN-2023-0006), and the Foundation for the Advancement of Theoretical Physics and Mathematics ``BASIS''.

\bibliography{biblio}
\bibliographystyle{apsrev4-1}



\begin{appendix}
\section{Appendix: Case of linearly polarized EM field}

Since the final formula describing the ratchet current in the case of linearly polarized EM wave is quite cumbersome to present in an explicit analytical form, we outline here the main stages of our calculation to ensure the reproducibility of our results (the continuation of Eq.~\eqref{current_lin})).
First, let us start with the calculation of current~(\ref{paramagnetic2}) by solving the system of Eqs.~(\ref{OP2}) with $n=1,2$. It is more convenient to separate two contributions, which appear after the substitution in $\Psi_1=\int d\textbf{r}_1dt_1\hat{G}(\textbf{rr}_1;tt_1)\hat{W}(\textbf{r}_1,t_1)\Psi_0(\textbf{r}_1,t_1)$ perturbation (a) $\hat{W}=\alpha \delta T_{c}\cos(\textbf{qr}_1)$ and (b) $\hat{W}=-2ie\gamma \varphi(\textbf{r}_1,t_1)$. 

Starting with the perturbation (a), we find (after performing the Fourier transformation):
\begin{widetext}
\begin{eqnarray}
&&\Psi_1(\textbf{r},t)=\frac{\alpha\delta T_c}{2}\sum_{\varepsilon,\textbf{p}}e^{i\textbf{p}\textbf{r}-i\varepsilon t}\left[\hat{G}_{\textbf{p}+\textbf{q}}(\varepsilon)e^{i\textbf{q}\textbf{r}}+\hat{G}_{\textbf{p}-\textbf{q}}(\varepsilon)e^{-i\textbf{qr}}\right]\Psi_0(\textbf{p},\varepsilon),\\&&\Psi_2(\textbf{r},t)=-2e^2\gamma^2h|\varphi|^2\sum_{\varepsilon,\textbf{p}}e^{i\textbf{pr}-i\varepsilon t}\left[e^{i\textbf{qr}}e^{i\phi}\hat{G}_{\textbf{p}+\textbf{q}}(\varepsilon)+e^{-i\textbf{qr}}e^{-i\phi}\hat{G}_{\textbf{p}-\textbf{q}}(\varepsilon)\right]\hat{G}_{\textbf{p}+\textbf{k}}(\varepsilon+\omega)\Psi_0(\textbf{p},\varepsilon)\\&&~~~~~-2e^2\gamma^2h|\varphi|^2\sum_{\varepsilon,\textbf{p}}e^{i\textbf{pr}-i\varepsilon t}\left[e^{i\textbf{qr}}e^{i\phi}\hat{G}_{\textbf{p}+\textbf{q}}(\varepsilon)\hat{G}_{\textbf{p}+\textbf{q}+\textbf{k}}(\varepsilon+\omega)+e^{-i\textbf{qr}}e^{-i\phi}\hat{G}_{\textbf{p}-\textbf{q}}(\varepsilon)\hat{G}_{\textbf{p}-\textbf{q}+\textbf{k}}(\varepsilon+\omega)\right]\Psi_0(\textbf{p},\varepsilon)\nonumber\\&&~~~~~+\{\omega,\textbf{k}\rightarrow -\omega,-\textbf{k}\}.\nonumber
\end{eqnarray}
 %
Substituting these expressions in Eq.~(\ref{paramagnetic2}) and averaging over the fluctuations and the spatial variable yields
%
\begin{eqnarray}\label{chto1}&&\overline{\langle\Psi_1^{\ast}(-i\nabla)\Psi_2\rangle}=-2\alpha T\delta T_ce^2h\gamma'\gamma^2|\varphi|^2\sum_{\varepsilon,\textbf{p}}|\hat{G}_{\textbf{p}}(\varepsilon)|^2\left\{\left[(\textbf{p}+\textbf{q})|\hat{G}_{\textbf{p}+\textbf{q}}(\varepsilon)|^2e^{i\phi}+(\textbf{p}-\textbf{q})|\hat{G}_{\textbf{p}-\textbf{q}}(\varepsilon)|^2e^{-i\phi}\right]\hat{G}_{\textbf{p+k}}(\varepsilon+\omega)\right.\nonumber\\
&&~~~~~~~~~~~~~~~+\left.\left[(\textbf{p}+\textbf{q})|\hat{G}_{\textbf{p}+\textbf{q}}(\varepsilon)|^2e^{i\phi}+(\textbf{p}-\textbf{q})|\hat{G}_{\textbf{p}-\textbf{q}}(\varepsilon)|^2e^{-i\phi}\right]\hat{G}_{\textbf{p+k+q}}(\varepsilon+\omega)\right\}+\{\omega,\textbf{k}\rightarrow -\omega,-\textbf{k}\}.
\end{eqnarray}
\end{widetext}
Since the ratchet effect should be $\propto \textbf{q}h$, we can simplify Eq.~(\ref{chto1}) by expanding it up to the first order with respect to $\textbf{q}$. This expansion allows us to combine the phase factors and pick out the common term $i\sin(\phi)$. 
Furthermore, the propagators can be expanded up to the second order with respect to $\textbf{k}$, more precisely $k_x^2$, $k_y^2$, and $k_xk_y$. 
Using the relation $\textbf{E}=-i\textbf{k}\varphi$ allows us to find the dependence of the current density on electric field projections, and the corresponding elements of the conductivity tensor $\chi_{\alpha\beta\gamma}$. 
Note, that Eq.~(\ref{chto1}) contains a complex number prefactor $i\gamma^2=i\gamma'^2-2\gamma'\gamma''$. Therefore, to achieve finite real-valued results, one should combine Eq.~(\ref{chto1}) with its complex conjugated counterpart $\overline{\langle\Psi_1(i\nabla)\Psi_2^{\ast}\rangle}$ and expand the propagators (where  necessarily) up to the first order in $\gamma''$. 
After the integration over the energy and polar angle, and summing up all the terms, we found that the transverse (to the direction of modulation of $T_c(\textbf{r})$) component of the current vanishes.

Next, we continue with the calculation of the current, which stems from the perturbation (b). 
The Fourier transform yields
\begin{widetext}
\begin{eqnarray}
&&\Psi_1(\textbf{r},t)=-2ie\gamma\sum_{\varepsilon,\textbf{p}}\varphi e^{i\textbf{pr}}e^{-i(\varepsilon+\omega) t}\left[\hat{G}_{\textbf{p}+\textbf{k}}(\varepsilon+\omega)+\frac{h}{2}e^{i\textbf{qr}+i\phi}\hat{G}_{\textbf{p}+\textbf{k}+\textbf{q}}(\varepsilon+\omega)+\frac{h}{2}e^{-i\textbf{qr}-i\phi}\hat{G}_{\textbf{p}+\textbf{k}+\textbf{q}}(\varepsilon+\omega)\right] \Psi_0(\textbf{p},\varepsilon)\nonumber\\&&~~~~~~~~~~~~~~~~~+\{\varphi,\omega,\textbf{k}\rightarrow \varphi^{\ast},-\omega,-\textbf{k}\},\\&&\Psi_2(\textbf{r},t)=-ie\gamma\alpha\delta T_c\sum_{\varepsilon,\textbf{p}}\varphi e^{i(\textbf{p}+\textbf{k})\textbf{r}-i(\varepsilon+\omega)t}\Psi_0(\textbf{r},t)\times\\&&~\left\{\left[\hat{G}_{\textbf{p}+\textbf{k}+\textbf{q}}(\varepsilon+\omega)e^{i\textbf{qr}}\left(\hat{G}_{\textbf{p}+\textbf{q}}(\varepsilon)+\hat{G}_{\textbf{p}+\textbf{k}}(\varepsilon+\omega)\right)+\hat{G}_{\textbf{p}+\textbf{k}-\textbf{q}}(\varepsilon+\omega)e^{-i\textbf{qr}}\left(\hat{G}_{\textbf{p}-\textbf{q}}(\varepsilon)+\hat{G}_{\textbf{p}+\textbf{k}}(\varepsilon+\omega)\right)\right]\right.+\nonumber\\&&~\left.\frac{ h}{2}\hat{G}_{\textbf{p}+\textbf{k}}(\varepsilon+\omega)\left[\hat{G}_{\textbf{p}-\textbf{q}}(\varepsilon)e^{i\phi}+\hat{G}_{\textbf{p}+\textbf{q}}(\varepsilon)e^{-i\phi}+\hat{G}_{\textbf{p}+\textbf{k}+\textbf{q}}(\varepsilon+\omega)e^{i\phi}+\hat{G}_{\textbf{p}+\textbf{k}-\textbf{q}}(\varepsilon+\omega)e^{-i\phi}\right]\right\}+\{\varphi,\omega,\textbf{k}\rightarrow \varphi^{\ast},-\omega,-\textbf{k}\}.\nonumber
\end{eqnarray}
%
Substituting these expressions in Eq.~(\ref{paramagnetic2}) and averaging over the fluctuations and the spatial coordinate gives
%
\begin{eqnarray}\label{chto2}
&&\overline{\langle\Psi_1^{\ast}(-i\nabla)\Psi_2\rangle}= 2\alpha T\delta T_c e^2 h |\gamma|^2\gamma'|\varphi|^2\sum_{\varepsilon,\textbf{p}}|\hat{G}_{\textbf{p}}(\varepsilon)|^2\left\{(\textbf{p}+\textbf{k})|\hat{G}_{\textbf{p+k}}(\varepsilon+\omega)|^2(2iq)\sin\phi\left[\left(\hat{G}_{\textbf{p+k}}(\varepsilon+\omega)\right)'_x-\left(\hat{G}_{\textbf{p}}(\varepsilon)\right)'_x\right]\right.\nonumber\\&&~~~~~~~~~~~~~~~+(\textbf{p}+\textbf{k}+\textbf{q})|\hat{G}_{\textbf{p}+\textbf{k}+\textbf{q}}(\varepsilon+\omega)|^2\left[\hat{G}_{\textbf{p}+\textbf{q}}(\varepsilon)+\hat{G}_{\textbf{p}+\textbf{k}}(\varepsilon+\omega)\right]e^{-i\phi}\nonumber\\&&~~~~~~~~~~~~~~~\left.+(\textbf{p}+\textbf{k}-\textbf{q})|\hat{G}_{\textbf{p}+\textbf{k}-\textbf{q}}(\varepsilon+\omega)|^2\left[\hat{G}_{\textbf{p}-\textbf{q}}(\varepsilon)+\hat{G}_{\textbf{p}+\textbf{k}}(\varepsilon+\omega)\right]e^{i\phi}\right\} +\{\omega,\textbf{k}\rightarrow -\omega,-\textbf{k}\}.
\end{eqnarray}   
\end{widetext}
As before, Eq.~(\ref{chto2}) is to be combined with its complex conjugate counter part $\overline{\langle\Psi_1(i\nabla)\Psi_2^{\ast}\rangle}$, expanded up to the first order with respect to $\textbf{q}$ and $\gamma''$, and up to the second order with respect to $\textbf{k}$. Again, terms that result in the $j_y$ component of the ratchet current ($j_y\perp q$) exactly vanish after summing them up. 
Therefore, Eqs.~(\ref{chto1}) and~(\ref{chto2}) contribute only to $j_x$ component of the ratchet current ($j_x\parallel q$). 

Finally, the paramagnetic current Eq.~(\ref{paramagnetic1}) also vanishes due to the nullification of the $\Psi_3(\textbf{r},t)$ correction.

\end{appendix}

\end{document}